**Dimerization-Induced Cross-Layer Quasi-Two-Dimensionality in Metallic Iridate $IrTe_2$**

G.L. Pascut,[1] K. Haule,[1] M.J. Gutmann,[2] S.A. Barnett,[3] A. Bombardi,[3] S. Artyukhin,[1] D. Vanderbilt,[1] J.J. Yang,[4] S.-W. Cheong,[1,4,5] V. Kiryukhin[1,5]

1. Department of Physics & Astronomy, Rutgers University, Piscataway, New Jersey 08854, USA
2. ISIS Facility, STFC-Rutherford Appleton Laboratory, Didcot OX11 OQX, UK
3. Diamond Light Source, Harwell Science and Innovation Campus, Didcot OX11 0DE, UK
4. Laboratory of Pohang Emergent Materials and Department of Physics, Pohang University of Science and Technology, Pohang 790-784, Korea
5. Rutgers Center for Emergent Materials, Rutgers University, Piscataway, New Jersey 08854, USA

**Abstract**

The crystal structure of layered metal $IrTe_2$ is determined using single-crystal x-ray diffraction. At *T*=220 K, it exhibits Ir and Te dimers forming a valence-bond crystal. Electronic structure calculations reveal an intriguing quasi-two-dimensional electronic state, with planes of reduced density of states cutting diagonally through the Ir and Te layers. These planes are formed by the Ir and Te dimers, which exhibit a signature of covalent bonding character development. Evidence for significant charge disproportionation among the dimerized and non-dimerized Ir (charge order) is also presented.

Compounds containing *5d* electrons have been the subject of numerous recent studies. Large spin-orbit coupling combined with electron-electron interactions gives rise to many intriguing phenomena, such as $J_{eff}$=1/2 Mott state,[1] correlated topological insulators,[2] charge ordering and ionic dimerization.[3] IrTe$_2$ is a layered chalcogenide metal composed of stacked layers of IrTe$_6$ octahedra forming a CdI$_2$-type structure[4] (space group *P-3m1*), see Fig. 1(a). Like many such chalcogenides, it exhibits a structural transition accompanied by a rise in the electrical resistivity, which is usually attributed to formation of a charge density wave (CDW) at low temperatures.[5] In IrTe$_2$, the structural modulation is characterized by the wave vector $q_0$=(1/5, 0, 1/5) with respect to the high-temperature Brillouin zone,[6] and the transition temperature is $T_S$~280 K.[7] Many recent experiments, however, are incompatible with the CDW nature of the modulated state in IrTe$_2$. For instance, a CDW gap is absent in the ARPES[8] and optical spectra,[9] and the structural modulation is highly non-sinusoidal.[7] A significant role of the orbital degrees of freedom has been discussed in many recent studies, and mechanisms based on orbital-driven Peierls instability have been proposed.[6,10] Ir *4f* core level x-ray photoemission experiments suggest a charge and orbital wave on the Ir sites.[11] Alternatively, a crystal field effect splitting the Te *p* orbitals was discussed as the driving force of the structural transition, based on optical spectroscopy experiments.[9] Depolymerization of the inter-layer Te bonds was also suggested as the origin of this transition.[7] The low-temperature state in IrTe$_2$ is currently under active investigation, and its nature is still being debated.

Knowledge of the crystallographic structure is necessary for understanding the modulated state in IrTe$_2$. X-ray powder diffraction for $T<T_S$ has been done, and an average monoclinic *C2/m* structure (with unresolved atomic positions),[5] as well as a 75-atom-large triclinic unit cell[12] were reported. However, a single-crystal diffraction experiment is needed to determine such a complex structure reliably. Herein, we present the structure of the modulated state at $T$=220 K determined by single-crystal x-ray diffraction. Both Ir and Te dimers form at this temperature. Electronic structure calculations provide evidence for covalent character development in the Ir dimers, as well as for charge disproportionation (charge order). The structural transition therefore appears to be driven by the energy gain due to Ir and Te dimerization, with Ir orbitals playing a key role. Most interestingly, the electronic structure of the modulated state is quasi-two-dimensional, with layers of reduced density of states at the Fermi level, formed by planes of Ir and Te dimers, cutting diagonally through the structural Ir and Te layers.

IrTe$_2$ single crystals were grown using Te flux, as described in Ref. [7]. X-ray diffraction measurements were done at room temperature and $T$=220 K using an Oxford Diffraction Supernova diffractometer equipped with a CCD detector and Mo K$_\alpha$ radiation. The room-temperature (RT) structure, shown in Fig. 1(a), is consistent with published data.[4] At $T$=220 K, our 102×33×17 μm IrTe$_2$ sample did not show any twinning (see supplementary Fig. S1), and all the observed ~23400 reflections were successfully indexed in the triclinic *P-1* cell shown in Fig. 1 (b) and described in Table I. This cell contains only 8 independent atoms, allowing for a reliable structural refinement. The structure was determined from 2523 unique reflections with $F^2>3\sigma$ using CrysAlisPro, Superflip, and JANA2006 software packages described in the Supplementary Material, the goodness of fit was $R_1$=0.057. The obtained atomic coordinates are given in Table I. Other details, including the thermal parameters, are given in the Supplementary Material.

The most remarkable feature of the *T*=220 K structure is Ir dimerization, see Fig 2(a). The Ir dimers form stripes in the triangular Ir layers. These stripes exhibit a staircase-like arrangement along the $c_0$ axis, resulting in a quintupling of the unit cell in the $a_0$ and $c_0$ directions of the high-temperature structure. The contraction of the Ir-Ir bonds in the dimers is striking – they are 20-23% shorter than all the other bonds in the Ir layer (3.119 Å *vs*. 3.905-4.030 Å). In Te triangular layers, Te-Te bonds above and below the Ir dimers contract by 10-17% compared to the other bonds (3.439 Å vs 3.807-4.114 Å), forming similar dimerized stripes, see Fig. 2(b). Both the Ir-Ir and Te-Te in-layer distances are 3.93 Å at RT. As at high temperature, the Te-Te bonds connecting the Te layers are shorter than the non-dimerized bonds in the layers due to Te polymerization. For *T*=220 K, they show less than 6% length variation ranging from 3.385-3.582 Å (the RT value is 3.498 Å). Finally, all the Ir-Te bonds are almost equal to each other and to the RT value, and vary by less than 2%.

Density functional theory (DFT) calculations were used to obtain information about structural energetics. They were performed using the full-potential (linearized) augmented plane-wave basis as implemented in WIEN2k code, and the GGA-PBE functional.[13] The DFT total-energy calculations show that both the high-*T* and low-*T* structures correspond to local minima separated by an energy barrier of the order of 40 meV per formula unit. To understand the nature of the dimerized state, we have carried out dynamical mean field theory (DMFT) calculations[14] in the charge self-consistent implementation[15] based on WIEN2k package.[16] The energy range in computing hybridization and self-energy spanned a 20eV window around the Fermi Energy ($E_F$), and corresponding values of local Coulomb repulsion $U$=4.5 eV and Hund's coupling $J$=0.8 eV were taken from Ref.[17]. The DMFT mass enhancement for Ir 5*d* (all $t_{2g}$) orbitals $m^*/m_{band}$ is between 1.15 and 1.2, hence only a minor mass enhancement due to electronic correlations was found. Spin-orbit coupling was fully included in both types of calculations.

Fig. 1 (c) shows the calculated Fermi Surface (FS) at RT; it is consistent with the published data.[9] For *T*=220 K, the system remains metallic with the FS shown in Fig. 1(d). The key property of this FS is its marked quasi-two-dimensional (2D) character. The direction normal to the electronic 2D planes is given by the reciprocal $c^*$ axis of the triclinic unit cell. This axis is normal to both $a$ and $b$ triclinic axes, and at an angle of 10° to the $c$ axis shown in Fig 1(b). The quasi-2D planes of the electronic structure cut diagonally through the structural Ir and Te planes.

The origin of this highly unusual electronic structure, which seemingly contradicts the structural motif of the crystal lattice, is illustrated in Fig. 3. It shows that the staircase-like arrangement of the Ir and Te dimers forms a 2D "wall" centered on Ir(3)-Ir(3) dimerized bonds. This wall cuts through the Ir and Te layers in the direction normal to c*. As shown in Fig 3 (a), the density of states (DOS) at $E_F$ is reduced dramatically for all the Ir and Te atoms within this wall. In contrast, only an insignificant DOS reduction is observed at $E_F$ for the Ir and Te atoms in the 2D planes away from the dimer walls (*eg.* Ir(1) and Te(2) sites). The largest DOS reduction occurs on the dimerized Ir sites. Consistently, the total DOS reduction at the $E_F$ that takes place at the structural transition is dominated by the Ir orbitals (about 2/3 of the reduction value), as shown in Fig. 4(a).

Dimerization of Ir(3) leads to significant changes in its *5d* DOS, see Fig. 4(b). The major effect is seen for the $d_{xy}$ orbitals that overlap directly in the Ir(3)-Ir(3) dimers. (Pseudo-cubic axes attached to $IrTe_6$ octahedra are used, see Fig. 4(b) for a sketch of the Ir(3) $d_{xy}$ orbitals.) Well-identified DOS peaks appear above and below $E_F$ for these orbitals. Such peaks are absent for the other Ir(3) *d* orbitals (and therefore less prominent in the Ir(3) total $5d\ t_{2g}$ DOS), for the other Ir atoms, and also at RT. Using chemistry language, they signify formation of bonding and anti-bonding molecular orbitals in the Ir dimers, see Fig. 4(b). These data, therefore, show evidence for covalent character development and bond formation in the Ir dimers. Ir dimerization is accompanied by a significant charge disproportion. At RT, the total *5d* orbital occupancy is calculated to be $n_d$=5.50, giving formal $Ir^{3.5+}$ valence that lies in between the $Ir^{4+}$ and $Ir^{3+}$ states discussed in the literature.[11,18] At $T$=220 K, $n_d$=5.22 ($Ir^{3.78+}$) for the dimerized Ir(3), while $n_d$=5.46 ($Ir^{3.54+}$) for the other Ir atoms. In a simplified ionic description, $Ir^{4+}$-like Ir(3) ions form strongly-bound dimers, reducing the DOS at $E_F$ and explaining the experimentally-observed drop in magnetic susceptibility,[5] while the other Ir sites remain $Ir^{3+}$-like and form a quasi-2D subsystem with higher electric conductivity.

As shown in Fig. 3(a), the Te ions also contribute to the DOS reduction at $E_F$, albeit not as significantly as Ir. The changes of the DOS of the Te orbitals at $T_S$ exhibit the same pattern as that shown by the Ir orbitals, see Fig. 4 (c). The DOS at $E_F$ is reduced significantly for the *4p* orbitals binding the Te(1)-Te(5) dimers in the Te planes, as well as for the *4p* orbitals in the shortest inter-plane Te pairs, Te(1)-Te(3). One could tentatively assign DOS peaks due to bonding and anti-bonding states, especially for the in-plane Te dimers, but such an interpretation is far less justified than that for the Ir(3) case. This possibly reflects the complex character of the network of Te bonds in the polymerized Te double planes, with different Te-Te bonds possessing different degrees of covalency. Outside the 2D walls of the dimerized Ir and Te, the Te $p$ orbitals largely preserve the RT character, as shown for the Te(2)-Te(2) interplane bonds in Fig. 4(c). Since the Ir(1) $5d$ orbitals also preserve the RT character, this explains the planes of large DOS at $E_F$ running through the network of Ir(1) and Te(2) atoms in the planes normal to $c^*$.

In conclusion, x-ray diffraction measurements combined with electronic structure calculations reveal an unusual quasi-2D electronic state in IrTe$_2$ at low temperatures. The planes of reduced DOS at the Fermi Energy consisting of dimerized Ir and Te ions cut diagonally through the triangular Ir and Te planes. The Ir dimers exhibit evidence for covalent bonding, as well as reduced occupancy of $5d$ orbitals (Ir$^{4+}$ character). The system can, therefore, be described using the valence-bond crystal language. Ir dimerization appears to be the main driving force for the structural transition at $T_S$. These results indicate that alternating layers with higher and lower electric conductivity should run in the direction normal to the triclinic c* axis in IrTe$_2$ at low temperatures. Confirmation of this prediction by either bulk (DC electrical conductivity), or local (scanning spectroscopy) measurements is, in our opinion, of significant interest.

We thank K.M. Rabe, Hongbin Zhang, C. Wilson, and T.J. Emge for fruitful discussions. The access to the x-ray diffraction facilities at the Research Complex at the Rutherford Appleton Laboratory is gratefully acknowledged. This work was supported by NSF DMREF Grant 12-33349. G.L.P. was supported by the NSF under Grant No. DMR-1004568. The work at Postech was supported by the Max Planck POSTECH/KOREA Research Initiative Program [Grant No. 2011-0031558] through NRF of Korea funded by MEST.

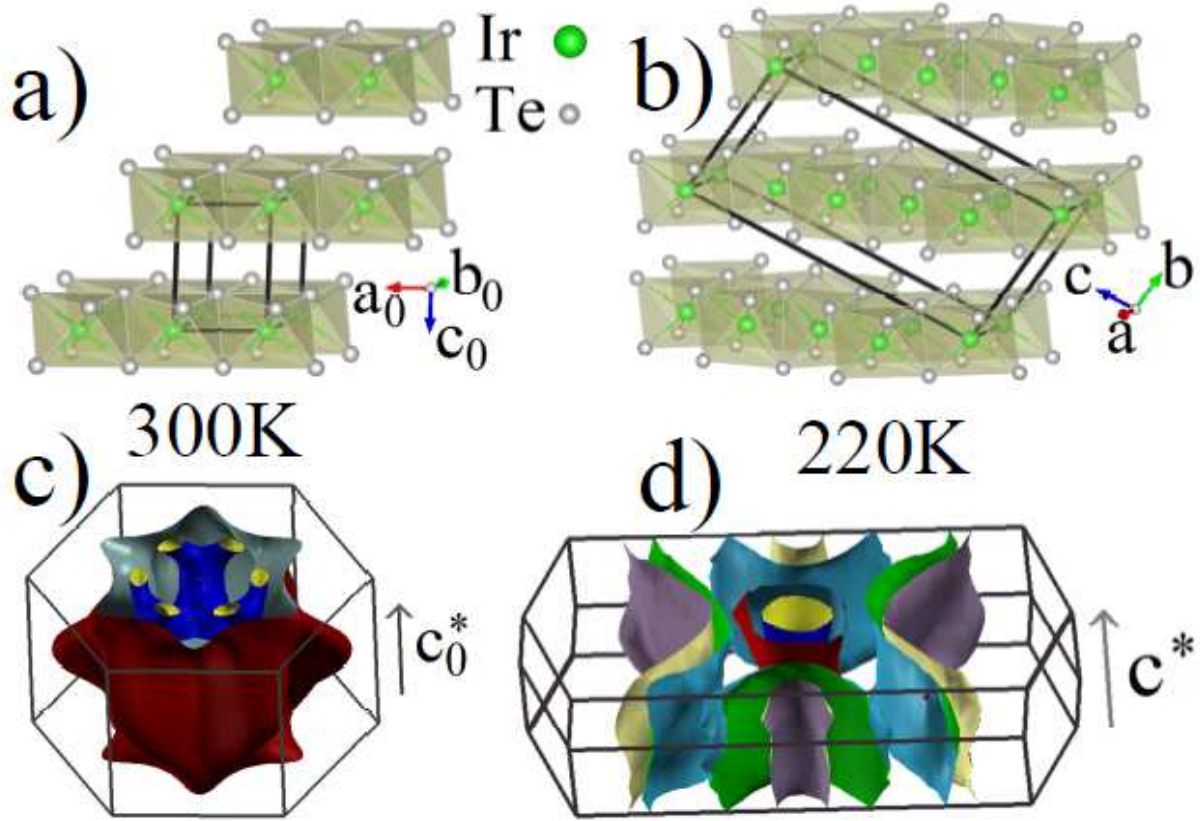

**Figure 1.** (Color online) The structure of IrTe$_2$ at room temperature (a), and $T$=220 K (b). The crystallographic unit cells (boxes) and axes (arrows) are shown. Calculated Fermi Surface at room temperature (c), and 220 K (d). The reciprocal axis $c_0{}^*$ runs along the $c_0$ crystallographic axis; $c^*$ is normal to the *ab* plane of the triclinic low-temperature unit cell.

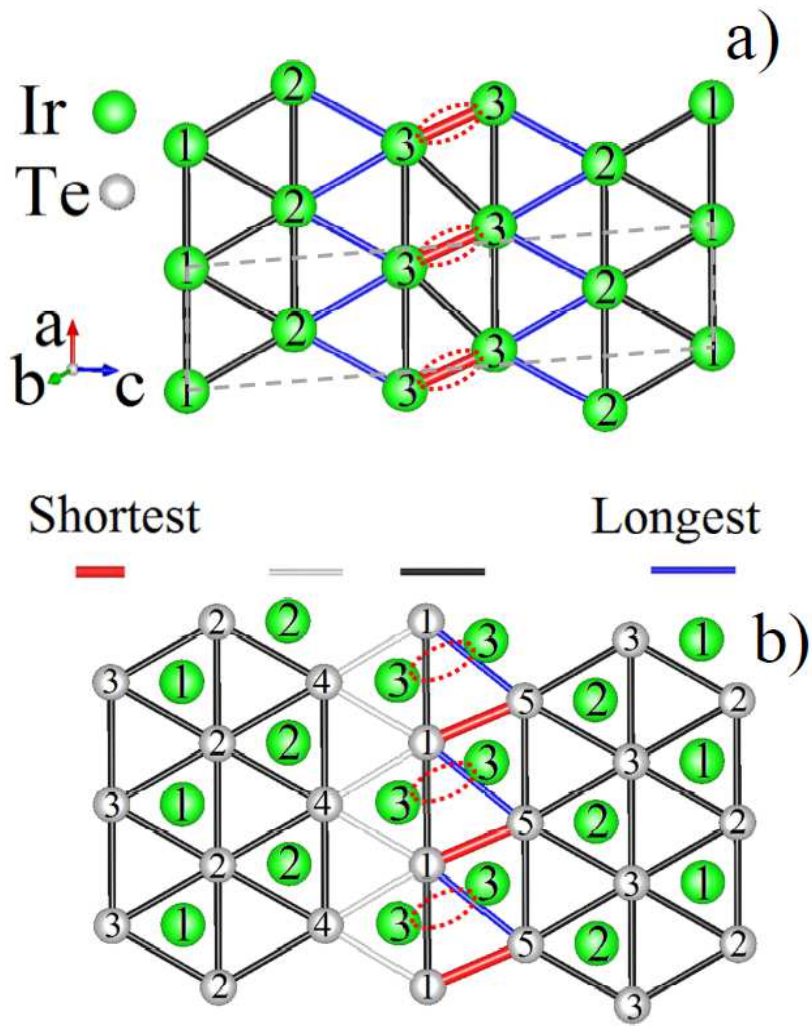

**Figure 2.** (Color online) (a) Triangular Ir layer at $T$=220 K. Dashed line shows the projection of the triclinic unit cell. (b) Triangular Te layer at $T$=220 K. The Ir layer directly above is also shown. In both panels, the bond lengths are color coded as shown in the legend, and Ir(3)-Ir(3) dimers are marked with dashed ovals.

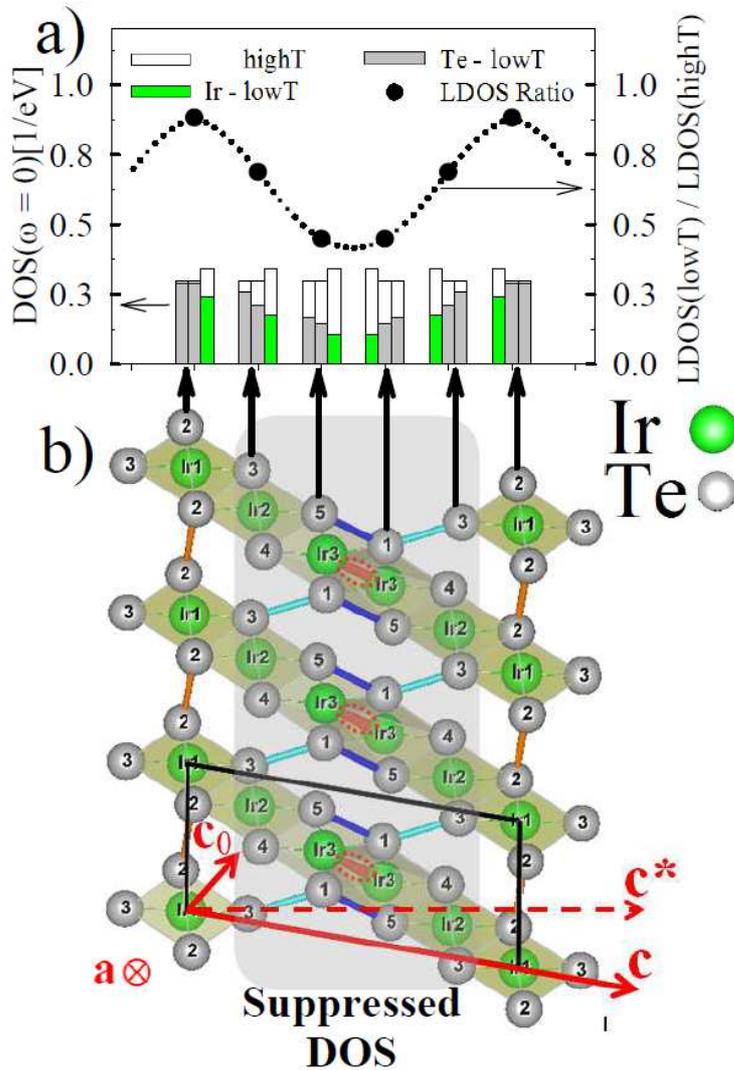

**Figure 3.** (Color) (a) The density of states at the $E_F$ for the planes of Ir and Te atoms running normal to the $c^*$ axis shown in panel (b) directly below. Each plane is identified with an arrow. The ratio of the low- (220 K) and high- (300 K) temperature DOS for each of these planes is also shown. (b) Projection of the $T$=220 K structure along the triclinic $a$ axis. The atoms form perfect columns, *i.e.* Ir(1) is directly below Ir(1), *etc*. Black box shows the projection of the unit cell. Ovals identify the Ir dimers. Shortened Te-Te bonds are shown in blue and cyan. The shaded region identifies a plane of suppressed DOS at $E_F$ formed by Ir and Te dimers. This plane is normal to the $c^*$ axis. In contrast, the structural Ir and Te layers run normal to the $c_0$ direction.

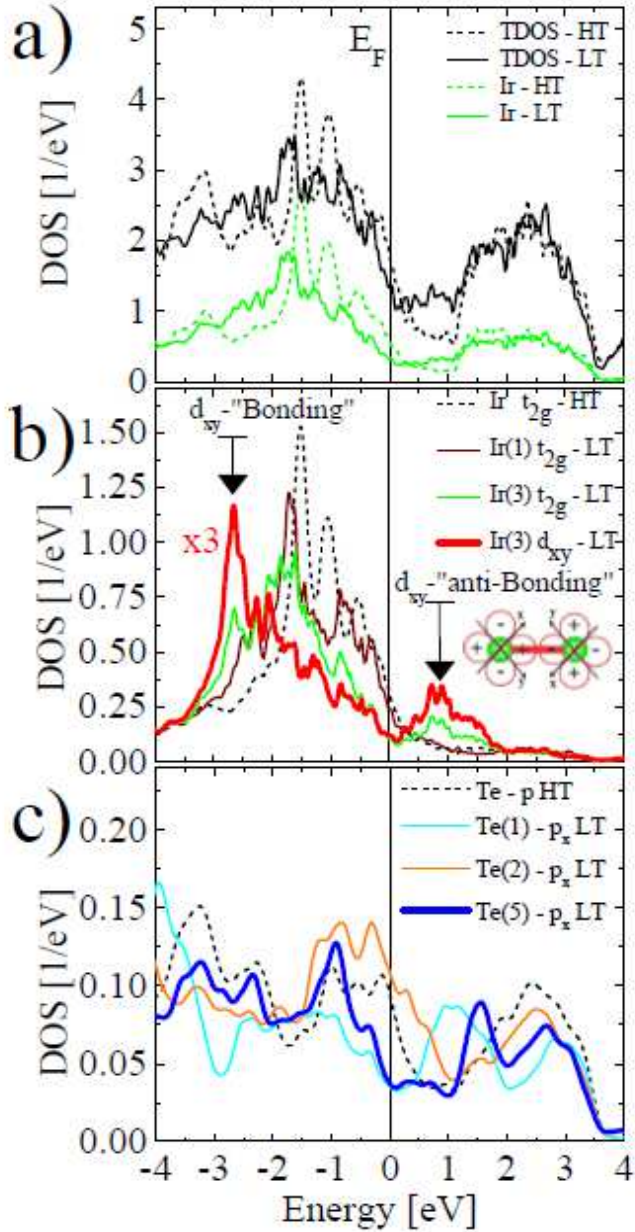

**Figure 4.** (Color) (a) Total density of states (TDOS), and Ir DOS at $T=300$ K (HT) and 220 K (LT). (b) Combined DOS of the $5d$ (all $t_{2g}$) orbitals for Ir(1) and Ir(3) at LT, as well as for any Ir at HT. The DOS for the Ir(3) $d_{xy}$ orbital at LT is also shown. The arrows point to the DOS peaks associated with the bonding and anti-bonding molecular orbitals in the Ir dimers. The corresponding anti-bonding orbital is sketched at right. The $x$, $y$, and $z$ axes for the $d$ orbitals are the local axes of the IrTe$_6$ octahedra. (c) LT DOS of the Te(1), Te(2), and Te(5) $4p_x$ orbitals for the Te(1)-Te(3), Te(2)-Te(2), and Te(5)-Te(1) bonds, respectively. The bonds are shown in Fig. 3(b) using the same color code. The $x$ axis of each $p_x$ orbital coincides with the corresponding Te-Te bond direction. The averaged HT Te $p$ DOS is also shown.

**Table I.** Atomic coordinates at *T*=220 K. The space group is *P-1*, the lattice parameters are $a$=3.9548(2) Å, $b$=6.6542(4) Å, $c$=14.4345(7) Å, $\alpha$=98.129(5)°, $\beta$=92.571(4)°, $\gamma$=107.119(5)°. The atomic positions are given in the lattice parameter units.

| Atom | x | y | z |
|---|---|---|---|
| Ir(1) | 0 | 0 | 0 |
| Ir(2) | 0.42570(12) | 0.78722(8) | 0.20327(3) |
| Ir(3) | 0.14156(12) | 0.42983(8) | 0.58879(3) |
| Te(1) | 0.21740(2) | 0.30048(13) | 0.41130(6) |
| Te(2) | 0.36780(2) | 0.72958(13) | 0.01680(5) |
| Te(3) | 0.05400(2) | 0.05471(13) | 0.18480(5) |
| Te(4) | 0.20320(2) | 0.47996(13) | 0.77765(5) |
| Te(5) | 0.51730(2) | 0.15875(13) | 0.61221(5) |